\documentstyle[12pt]{article}
\begin{document}
\begin{center}
\title{Covariant quark model of form factors \\ in the heavy mass limit}
\author{ A. Le Yaouanc, L. Oliver, O. P\`ene and J.-C. Raynal} \par
\maketitle
{Laboratoire de Physique Th\'eorique et Hautes
Energies\footnote{Laboratoire
associ\'e au
Centre National de la Recherche Scientifique - URA D00063}}\\
{Universit\'e de Paris XI, B\^atiment 211, 91405 Orsay Cedex,
France}\\
\end{center}

\vskip 5 mm
\begin{flushright}
LPTHE-Orsay 95/52,\\
hep-ph/9507342
\end{flushright}
\noindent {\bf Abstract}\\
\begin{abstract}
 We show that quark models of current matrix-elements based on the
Bakamjian-Thomas construction of relativistic states with a fixed number of
particles, plus the additivity assumption, are covariant in the heavy-quark
limit and satisfy the full set of heavy-quark symmetry relations discovered
by Isgur and Wise. We find the lower bound of $\rho^2$ in such models to be
$3/4$ for ground state mesons, independently of any parameter. Another welcome
property
of these models is that in the infinite momentum limit the wave functions
vanish outside the domain $0\le x \le 1$.
\end{abstract}
\newpage

\section{Introduction}

The necessity of a relativistic treatment of hadron center-of-mass motion in
quark models is
manifest in the calculation of form factors at high three-dimensional momentum
transfer  $\vec q$  and
effort in this direction have been made since a long time. On the other hand,
it has been realized
rather recently that in QCD hadronic form factors must satisfy a set of
remarkable relations in the
limit where the mass of the active quark in the hadron is made heavy
\cite{heavy quarks}, the
so-called heavy quark symmetries.

In the past few years ago \cite{nous1}, we have noticed that models based on a
very simple treatment of hadron
motion with Lorentz boost of spins and Lorentz contraction of spatial wave
functions - which we have
formulated in the 70's \cite{nous2} - indeed present these symmetry properties
in the heavy quark limit \cite{nous1}, \cite{nous3}.
This is not at all trivial because, as can be observed in the literature, most
current models do not satisfy these properties \cite{aleksan} (or enforce them
by hand). However our own models \cite{nous1} \cite{nous2} \cite{nous3} have
serious drawbacks.
They are not covariant and they use, in addition to the basic assumptions of
quark models, a series
of approximations which are not settled in a well-defined framework. In
addition, they do not show the expected behavior at large $\vec q\,^2$, which
in turn is related to the fact that the null-plane limit of the wave functions
do not vanish outside the domain $0\le x \le 1$.

We will use here such a well-defined framework which in addition turns out to
have the outstanding
merit of being covariant in the heavy quark limit and maintain the scaling
properties found in the
na\"\i ve models. The purpose of this letter is to show these properties of
this new approach and to
derive a bound on the widely discussed parameter $\rho^2$, the slope of the
Isgur-Wise function. In addition, the class of models obtained in that way
solves the above-mentioned problem at large $\vec q\,^2$, since, in the
infinite momentum limit, the wave functions vanish outside the domain $0\le x
\le 1$, as will be proved elsewhere.

Let us say, before proceeding, a few general words on the method. Although this
does not seem to be
expected from the knowledge of field theory, it is a very old finding
\cite{BT},\cite{osborn},\cite{keister} that
one can make an important step towards a fully relativistic theory with a fixed
number of interacting
constituents, with wave functions implementing a representation of the
Poincar\'e group through the
construction of the full set of generators, and with a rest-frame Hamiltonian
(or mass operator)
containing rather standard (non-relativistic looking) potentials depending on
relative coordinates,
two-dimensional Pauli spins, \ldots

In this framework, the problem of knowing the relativistic wave function in
motion in terms of the
wave function at rest is solved exactly and in a rather simple manner, through
a change of variables
which is known once for all and does not depend on the interaction.

The whole solution relies on a complete separation between two types of
variables, related to
individual particle variables by explicit expressions which are simple at least
in momentum space.
On the one hand, we have global variables which describe the whole system in
analogy with the
one-particle state. On the other hand, we have internal variables which
somewhat generalise the
relative variables of non-relativistic systems, a major property being that the
two types of
variables are commuting as operators.

The important point is then that one can construct the Poincar\'e generators as
the ones of a free
particle described by the global variables (total momentum  $\vec P$  and
relativistic center-of-mass
position  $\vec R$)  and a mass which can be chosen arbitrarily provided it
depends only on the
internal variables.

This framework would be ideally suited to formulate relativistic quark models
except for one serious
drawback. There is no known {\em covariant\/} current operator, and in
particular the usual one-quark
free current, which corresponds to the basic {\it additivity assumption} of
quarks models, is not covariant {\it in general}
when sandwiched between hadron states. This failure entails that we have not a
satisfactory
relativistic model for {\em transitions}, which was however precisely the
initial motivation for
appealing to this treatment. We have then to return in general once more to the
old discussions on a
possible approximate covariance under certain particular conditions, or on the
choice of a best
reference frame.

One remarkable exception to this failure is the case of the {\em heavy quark
limit}. Let us recall
that this limit consists in considering systems containing one quark very heavy
with respect to all
the others, and moreover to consider transitions between such ``heavy-light''
hadrons, where it is the
heavy quark which endows electroweak interaction through some external current.
In this limit, we
find that the above framework with the one-quark current gives a {\em
covariant\/} model for
transitions. It is then on this limit that we shall concentrate after a
presentation of the general
formalism, leaving the study of other situations for future discussions. We
recall that the interest
of this limit is not purely theoretical. It is believed to be roughly realized
in the  $B \to D^{(*)} l
\nu$  semi-leptonic decays, where one is measuring in particular with
increasing accuracy the slope
$\rho^2$ at the no-recoil point,  and where the knowledge of form factors is
expected to improve much in
the future. It is interesting that our lower bound on  $\rho^2$  is rather
close to the value estimated
by CLEO II experiment.

On the theoretical side, we must emphasize that the model presented in this
letter is not just one
more quark model. It has the interest of embedding many recent attempts.
Indeed, we find that
certain recent models for mesons or baryons are explicitly based on the B-T
formalism plus the free
quark current \cite{keister},\cite{szczepaniak}. On the other hand, there are
many approaches directly formulated on the null-plane,  among
which several \cite{terent'ev},\cite{jaus}, \cite{cardarelli}, \cite{capstick},
can be shown to be the $P = \infty$  limit of the present approach. They
will have the same heavy quark limit obtained in this letter, since we show
this limit to be
covariant.  Also, according to our findings, it
seems that the intuitive approach of Close and Wambach \cite{close}, directly
formulated for the
heavy quark limit, leads to the same final expressions. One consequence is that
the lower bound
on
$\rho^2$  we have found is of general interest in that it will apply to a large
class of
proposed models.

Of course, this does not apply to all models encountered in the literature, and
one may find
seemingly similar approaches which differ by certain details which, in the end,
reveal crucial
\cite{dziembovski}. We have also definitely
different approaches leading to covariant and
scaling models. like the one of Kaidalov \cite{kaidalov}.

As to our previous model \cite{nous1}, it comes out that it could be defined,
at the start, as an approximation to
the present one, which would consist mainly in neglecting internal momenta with
respect to the quark
masses. However, although this could seem quite natural in a quark model, it is
found to lead
finally to drastic differences with the present one, especially at large
${\vec q}\,^2$,  on which
we will comment elsewhere.

\section{Relativistic quark models for currents in the Bakamjian-Thomas
formalism.}

In the present section, we will not yet take the infinite mass limit.
The $n$-particle Hilbert space, which is naturally the tensor product of the
$n$  individual
one-particle Hilbert spaces, is made of functions  $\Psi_{s_1,\ldots s_n}(\vec
p_1, \ldots\vec
p_n)$  of the so-called one-particle variables, spins  $\vec S_i$  and momenta
$\vec p_i$.  Assuming
that particle $1$ is the active particle, the additivity hypothesis means that
the current density
operator in the $n$-particle Hilbert space will be the tensor product of the
current density operator
on particle $1$ by the identity operators on particles $2, \ldots n$.  This
writes :
\begin{eqnarray}
		\lefteqn{\langle\Psi'| O |\Psi\rangle  =  \int \frac{d{\vec
p}^{\,\prime}_1}{{(2\pi)}^3}
\frac{d\vec p_1}{{(2\pi)}^3}\,  (\prod_{i=2}^n\frac{d\vec p_i}{{(2\pi)}^3}) \;
\sum_{s'_1,s_1} \;
\sum_{s_2,\ldots s_n}}\\
			&&\Psi'_{s'_1,s_2,\ldots s_n}({\vec p}^{\,\prime}_1, \vec p_2,\ldots\vec
p_n)^{\displaystyle*}
O({\vec p}^{\,\prime}_1,\vec p_1)_{s'_1,s_1}  \Psi_{s_1,s_2,\ldots s_n}(\vec
p_1, \vec p_2,\ldots\vec
p_n)\nonumber
\end{eqnarray}
where  from now on the primes will denote the final states, and $O({\vec
p}^{\,\prime},\vec p)_{s',s}$  is the matrix element between one-particle
states:
\begin{equation}
		O({\vec p}^{\,\prime},\vec p)_{s',s}  =  \langle {\vec p}^{\,\prime}, s' | O
| \vec p, s \rangle
\end{equation}
The one-particle states that we use are defined, including their normalisation,
by eq. (6) below.

Let us describe the essentials (for our purpose) of the B-T model, which is a
way to implement exact
Poincar\'e group transformations for a finite, fixed, number of interacting
particles. In order to
define the Poincar\'e transformations, one introduces another set of variables,
namely, total
momentum  $\vec P$,  internal momenta  $\vec k_1, \ldots\vec k_n$  ($\sum \vec
k_i = 0$)  and
internal spins $\vec S'_i$. The {\it unitary} transformation which relates the
previous wave
functions to the wave functions  $\Psi^{int}_{s_1,\ldots s_n}(\vec P, \vec
k_2,\ldots\vec k_n)$, depending on
the internal variables, is the following:
\begin{eqnarray}
	\Psi_{s_1,\ldots s_n}(\vec p_1,\ldots\vec p_n)  &=& \sqrt{\frac{\Sigma
p_j^0}{M_0}} \;
(\prod_{i=1}^n\frac{\sqrt{k^0_i}}{\sqrt{p^0_i}})\sum_{s'_1,\ldots s'_n}\\
				&&\qquad	  (\prod_{i=1}^n D_i({\mbox{\boldmath $R$}}_i)_{s_i,s'_i})\;
\Psi^{int}_{s'_1,\ldots s'_n}(\Sigma \vec p_i, \vec k_2,\ldots\vec
k_n)\nonumber
\end{eqnarray}
where (on the right), the vectors  $\vec k_i$,  the  $0$-components  $k_i^0$
and
$p_i^0$,  $M_0$,  and the Wigner rotations  $\mbox{\boldmath $R$}_i$  are
functions of the  $\vec p_i$  defined as
follows :
\begin{eqnarray}
		\lefteqn{p_i^0 = \sqrt{\vec p_i^{\,2}+m_i^2},}\hspace{1.5in} &&		M_0  =
\sqrt{(\Sigma p_j)^2},\\
		\lefteqn{k_i  =  \mbox{\boldmath $B$}^{-1}_{\Sigma p_j} p_i,} \hspace{1.5in}
&&	\mbox{\boldmath $R$}_i  =  \mbox{\boldmath
$B$}^{-1}_{p_i} \mbox{\boldmath $B$}_{\Sigma p_j} \mbox{\boldmath $B$}_{k_i}
\nonumber
\end{eqnarray}
(notations :  $\mbox{\boldmath $B$}_p$  is the boost  $(\sqrt{p^2}, \vec0) \to
p$,  $D_i(\mbox{\boldmath $R$})$ is the matrix of
the rotation  {\boldmath $R$}  for the spin  $S_i$). Let us stress some virtues
of unitarity: starting from an orthonormal set of internal wave functions, one
gets an orthonormal set of wave functions in any frame.

The Poincar\'e generators are then defined as  $\vec P$  for the space
translations and,
\begin{eqnarray}
		&&H  =  P^0  =  \sqrt{\vec P^2+M^2}\\
		&&\vec J  =  - i \vec P \times \frac\partial{\partial\vec P}  +  \vec
S,\qquad\qquad	\vec S  =
\sum_{i=1}^n\vec S'_i - i\sum_{i=2}^n \vec k_i \times
\frac\partial{\partial\vec k_i} \nonumber \\
		&&\vec K  =  - \frac i2  [P^0, \frac\partial{\partial\vec P}]_+  - \frac{\vec
P\times\vec
S}{P^0+M}\nonumber
\end{eqnarray}
where, in order to satisfy the Poincar\'e commutators, the sole requirement is
that the mass
operator  $M$  depends only on the internal variables and is invariant by
rotation. Namely,  $M$  must
commute with  $\vec P$,  $\frac\partial{\partial\vec P}$  and  $\vec S$. It
is important to realize that, in the interacting case, we have to deal
simultaneously with two different mass operators, $M_0$ of equation (4), and
the true mass operator $M$ which appears in (5) and contains the interaction.

The non-interacting case corresponds to   $M = M_0 = \sum_{i=1}^n\sqrt{\vec
k_i^2+m_i^2}$  and,
through the transformation (3), the generators (5) reduce then to the sum of
the one-particle free
generators, which is the main virtue of (3).

It is useful to notice that, if  $M$  stands for the mass of a particle and
$\vec S$  for its spin,
formulae  (5)  gives precisely the one-particle free generators when (as
assumed here) the moving
spin states are defined by
\begin{eqnarray}
		&&|\vec P, \mu \rangle  =  \sqrt{\frac M{P_0}}  \mbox{\boldmath $B$}_P
|\vec0, \mu \rangle\\
		&&(P^0 = \sqrt{\vec P^2+M^2},\qquad     \langle\vec P', \mu' |\vec P, \mu
\rangle  =  (2\pi)^3\:
\delta(\vec P'-\vec P)\: \delta_{\mu',\mu})\nonumber
\end{eqnarray}
The generators follows from the finite transformations, which are given by :
\begin{equation}
		\Lambda |\vec P, \mu \rangle  =  \sqrt{\frac{(\Lambda P)^0}{P_0}}  \sum_{\mu}
D(\mbox{\boldmath $B$}_{\Lambda P}^{-1}\Lambda\mbox{\boldmath
$B$}_P)_{\mu',\mu}\; |\vec{\Lambda P},\mu' \rangle
\end{equation}
where  $D(\mbox{\boldmath $R$})$  is the matrix of the rotation  {\boldmath
$R$}  for the spin  $S$.

First it is then easy to see why  (5)  satisfy the Poincar\'e Lie algebra,
since the calculation of
the commutators in (5) is the same as in the one-particle free case, due to the
commutativity of the
mass operator  $M$  with  $\vec P$,  $\frac\partial{\partial\vec P}$  and
$\vec S$.

Next the finite transformations generated by (5)  are just given (on
eigenstates of  $M$  and
$\vec P$) by (7), except that, instead of acting on the spin  $\vec S$,  the
Wigner rotation
$\mbox{\boldmath $B$}_{\Lambda P}^{-1}\Lambda\mbox{\boldmath $B$}_P$  applies
now to the internal spins  $\vec S'_i$  and
the internal momenta $\vec k_i$.  In fact they are directly given by (7) on
eigenstates of  $M$,
$\vec P$, $S^2$, $S_z$.

We are now in position to construct the wave functions of moving bound states.
Let
$\varphi_{s_1,\ldots s_n}(\vec k_2,\ldots\vec k_n)$  be an eigenstate of  $M$,
$S^2$, $S_z$  (in the
Hilbert space reduced with respect to  $\vec P$). The corresponding
(generalized) eigenstate of
$M$,
$\vec P$, $S^2$, $S_z$  with  $\vec P = 0$  writes :
\begin{equation}
		\Psi^{(\vec P=\vec0),int}_{s_1,\ldots s_n}(\vec Q, \vec k_2,\ldots\vec k_n)
=  (2\pi)^3\,
\delta(\vec Q)\;  \varphi_{s_1,\ldots s_n}(\vec k_2,\ldots\vec k_n)
\end{equation}
The moving wave function  $\Psi^{(\vec P),int}$  is obtained by applying the
boost  $\mbox{\boldmath $B$}_P$
($P^0 =\sqrt{\vec P^2+M^2}$)  to the wave function  (8).  Since we start from
$\vec P = 0$,  we may
use eq. (6), and the result is simply :
\begin{equation}
		\Psi^{(\vec P),int}_{s_1,\ldots s_n}(\vec Q, \vec k_2,\ldots\vec k_n)=
(2\pi)^3\,
\delta(\vec Q-\vec P)\;  \varphi_{s_1,\ldots s_n}(\vec k_2,\ldots\vec k_n)
\end{equation}
The wave function in the one-particle variables, for a bound state of momentum
$\vec P$,  is then
obtained from (3) :
\begin{eqnarray}
		\lefteqn{\Psi^{(\vec P)}_{s_1,\ldots s_n}(\vec p_1,\ldots\vec p_n)  =
\sqrt{\frac{\Sigma p_j^0}{M_0}}\:
(\prod_{i=1}^n\frac{\sqrt{k^0_i}}{\sqrt{p^0_i}})}\qquad\\
				&&\sum_{s'_1,\ldots s'_n}\;(\prod_{i=1}^n D_i(\mbox{\boldmath
$R$}_i)_{s_i,s'_i})\;(2\pi)^3\,
\delta(\Sigma\vec p_i-\vec P)\;\varphi_{s'_1,\ldots s'_n}(\vec k_2,\ldots\vec
k_n)\nonumber
\end{eqnarray}
When  $\vec P = 0$,  this reduces to :
\begin{equation}
		\Psi^{(\vec0)}_{s_1,\ldots s_n}(\vec p_1,\ldots\vec p_n)  =   (2\pi)^3\,
\delta(\Sigma\vec p_i)\;
\varphi_{s_1,\ldots s_n}(\vec p_2,\ldots\vec p_n)
\end{equation}
showing that  $\varphi$  is just the rest-frame internal bound state wave
function.

Introducing the momentum eigenstates given by (10) in formula (1), one gets:
\begin{eqnarray}
	\lefteqn{\langle\vec P'| O |\vec P\rangle  =  \int (\prod_{i=2}^n \frac{d\vec
p_i}{(2\pi)^3})\;
\sqrt{\frac{\Sigma p_j^{\prime0}\Sigma p_j^0}{M'_0M_0}}\;  (\prod_{i=1}^n
\frac{\sqrt{k^{\prime0}_ik^0_i}}{\sqrt{p^{\prime0}_ip^0_i}})}\qquad\\
		&&\sum_{s'_1,\ldots s'_n}\;  \sum_{s_1,\ldots s_n}
\varphi'_{s'_1,\ldots s'_n}(\vec k'_2,\ldots\vec
k'_n)^{\displaystyle*}\nonumber\\
		&&[D'_1(\mbox{\boldmath $R$}_1'^{-1})O(\vec p^{\,\prime}_1,\vec
p_1)D_1(\mbox{\boldmath $R$}_1)]_{s'_1,s_1}\;  [\prod_{i=2}^n
D_i(\mbox{\boldmath $R$}_i'^{-1}\mbox{\boldmath $R$}_i)_{s'_i,s_i}]\;
\varphi_{s_1,\ldots s_n}(\vec k_2,\ldots\vec k_n)
\nonumber
\end{eqnarray}
where (on the right) the quantities  $k_i$,  $p_i^0$,  $M^0$,  $\mbox{\boldmath
$R$}_i$  are the functions of
$\vec p_2\ldots\vec p_n$  given by (4) with  $\vec p_1  =  \vec P -
\sum_{i=2}^n\vec p_i$ ,  and the
analogous primed quantities are also given by (4) replacing  $\vec p_1$  by
$\vec p^{\,\prime}_1
= \vec P' - \sum_{i=2}^n \vec p_i$  and  $m_1$  by  $m'_1$.

As said before, eq. (12) does not give a in general a covariant model for the
current matrix elements. The reason
is that the one-particle operator  $O$  is covariant with respect to the free
one-particle Lorentz
transformations, while the transformations (5) depend on the interaction.
However, it turns out that
(12) {\em becomes covariant in the heavy mass limit\/}  $m_1, m'_1 \to \infty$.
 This is a general
result which requires only the covariance of  $O$  for free one-particle
transformations. We leave
the proof to a later publication and restrict in this letter to the case of
mesons, namely systems of
two spin $1/2$ particles. Furthermore, in accordance with heavy mass limit
ideas in QCD, we assume
spin-independent interaction, and satisfy in this case Isgur-Wise scaling. We
consider only the
ground states, pseudoscalar and vector mesons.

\section{Formulation in terms of Dirac matrices.}

Under the assumption of spin independent forces, the pseudoscalar and vector
wave functions write
\begin{eqnarray}
	&&\varphi_{s_1,s_2}(\vec k_2)  =  \frac i{\sqrt2}\; (\sigma_2)_{s_1,s_2}\;
\varphi(\vec k_2)\\
	&&\varphi^{(\vec e)}_{s_1,s_2}(\vec k_2)  =  \frac i{\sqrt2}\; ((\vec
e.\vec\sigma)\sigma_2)_{s_1,s_2}\; \varphi(\vec k_2) \nonumber
\end{eqnarray}
where  $\varphi(\vec k_2)$  is only required to be invariant by rotation, and
$\vec e$  is the rest
frame polarisation vector of the vector meson. The spin sums in eq. (12) reduce
to the trace of a
$2\times2$  matrix. For example, an harmonic oscillator potential leads to:
\[
\varphi(\vec k_2)= (2\pi)^{\frac32}\left({R/ \sqrt{\pi} } \right)^{\frac 3 2}
e^{-R^2\vec k_2^2 /2}\]

Using the relation  $\sigma_2D(\mbox{\boldmath $R$})\sigma_2  =
D(\mbox{\boldmath $R$}^{-1})^t$,  one gets:
\begin{eqnarray}
	\lefteqn{\langle\vec P', \vec e^{\,\prime} | O |\vec P, \vec e \rangle  =
\int \frac{d\vec
p_2}{(2\pi)^3}  \sqrt{\frac{\Sigma p_j'^0\Sigma p_j^0}{M'_0M_0}}
\frac{\sqrt{k'^0_1k^0_1}}{\sqrt{p'^0_1p^0_1}}
\frac{\sqrt{k'^0_2k^0_2}}{p^0_2}}\\
		&&\frac12\; \mathrm { Tr}[(\vec e^{\,\prime}.\vec\sigma)^\dagger
D(\mbox{\boldmath $R$}_1'^{-1})O(\vec
p^{\,\prime}_1,\vec p_1)D(\mbox{\boldmath $R$}_1)(\vec
e.\vec\sigma)D(\mbox{\boldmath $R$}_2^{-1}\mbox{\boldmath $R$}'_2)]
\;\varphi'(\vec k'_2)^{\displaystyle*} \varphi(\vec k_2) \nonumber
\end{eqnarray}
for the matrix element between vector mesons. The other matrix elements are
obtained by omitting
$(\vec e.\vec\sigma)$  or  $(\vec e^{\,\prime}.\vec\sigma)$  or both under the
trace.

Next, this formula can be written in a more familiar form, involving the
$4\times4$  Dirac matrices
instead of the  $2\times2$  Pauli matrices. Between vector mesons, one gets
\begin{eqnarray}
	\lefteqn{\langle\vec P', \epsilon' | O |\vec P, \epsilon \rangle  =  \int
\frac{d\vec p_2}{(2\pi)^3}  \frac1{p^0_2}  F(\vec p_2, \vec P', \vec P)}\\
		 &&\frac1{16}
\mathrm { Tr}[O(m_1+\rlap/p_1)(1+\rlap/u)\gamma_5\rlap/\epsilon_u
(m_2+\rlap/p_2)\gamma_5
\rlap/\epsilon'^{\displaystyle*}_{u'}(1+\rlap/u')(m'_1+\rlap/p'_1)]\;
\varphi'(\vec
k'_2)^{\displaystyle*} \varphi(\vec k_2) \nonumber
\end{eqnarray}
\begin{equation}
	F(\vec p_2, \vec P', \vec P)  =   \frac{\sqrt{u'^0u^0}}{p'^0_1p^0_1}
\frac{\sqrt{k_1'^0}}{\sqrt{k_1'^0+m'_1}} \frac{\sqrt{k_1^0}}{\sqrt{k_1^0+m_1}}
\frac{\sqrt{k_2'^0}}{\sqrt{k_2'^0+m_2}} \frac{\sqrt{k_2^0}}{\sqrt{k_2^0+m_2}}
\end{equation}
In (15), the following additional notations are used.  The unit 4-vectors  $u$
and  $u'$ :
\begin{equation}
		u  =  \frac{p_1+p_2}{M_0},\qquad\qquad			u'  =  \frac{p'_1+p_2}{M'_0}
\end{equation}
The 4-vectors  $\epsilon_u$  and  $\epsilon'_{u'}$  are related to the
polarisation 4-vectors
$\epsilon = \mbox{\boldmath $B$}_P (0,\vec e)$  and  $\epsilon' =
\mbox{\boldmath $B$}_{P'} (0, \vec e^{\,\prime})$  by
\begin{equation}
		\epsilon_u  =  \mbox{\boldmath $B$}_u \mbox{\boldmath $B$}_P^{-1} \epsilon,
\qquad\qquad	\epsilon'_{u'} =  \mbox{\boldmath $B$}_{u'}
\mbox{\boldmath $B$}_{P'}^{-1} \epsilon'
\end{equation}
Moreover, the  $O$  under the trace stands for the Dirac matrix appropriate to
the current
considered, for example $\gamma_\mu$, $\gamma_\mu\gamma_5$, etc. The matrix
elements for other mesons are obtained by omitting
$\gamma_5\rlap/\epsilon_u$  or  $\gamma_5\rlap/\epsilon'^{\displaystyle*}_{u'}$
 or both under the
trace.

Let us describe the main steps to deduce  (15, 16) from (14). $O(\vec
p^{\,\prime},\vec p)_{s',s}$
is the matrix element of  $\gamma^0O$  between spinors of the form
\begin{equation}
		\sqrt{\frac m{p^0}\;} \mbox{\boldmath $B$}_p \pmatrix{\chi_s\cr0}  =
\sqrt{\frac{p^0+m}{2p^0}}
\pmatrix{\chi_s\cr\cr\displaystyle{\frac{\vec p.\vec\sigma}{p^0{+}m} \chi_s}}
\end{equation}
where the boost  $\mbox{\boldmath $B$}_p$  also stands for its matrix in the
Dirac representation :
\begin{equation}
		\mbox{\boldmath $B$}_p  =   \frac{m + \rlap/p\gamma^0}{\sqrt{2m(p^0+m)}}
\end{equation}
The  $2\times2$  Pauli matrix in (14) may be considered as a  $4\times4$  Dirac
matrix with only its
upper left  $2\times2$  block non vanishing. The matrix  $O(\vec
p^{\,\prime},\vec p)$  then writes :
\begin{equation}
		O(\vec p^{\,\prime}_1,\vec p_1)  =
\frac{\sqrt{m'_1m_1}}{\sqrt{p'^0_1p^0_1}}\;
\frac{1+\gamma^0}2 \mbox{\boldmath $B$}_{p'_1}^{-1}O\mbox{\boldmath $B$}_{p_1}
\frac{1+\gamma^0}2
\end{equation}
(using  $\gamma^0\mbox{\boldmath $B$}_p\gamma^0 = \mbox{\boldmath
$B$}_p^{-1}$).  Also we replace the Wigner rotations in
(14) by their expressions (4)  as products of three boosts. The resulting
expressions contain
matrices sandwiched between  $\mbox{\boldmath $B$}_u$  and  $\mbox{\boldmath
$B$}^{-1}_u$  and between  $\mbox{\boldmath $B$}_{u'}$  and
$\mbox{\boldmath $B$}^{-1}_{u'}$.  They are reduced using
\begin{equation}
		\mbox{\boldmath $B$}_u(1+\gamma^0)\mbox{\boldmath
$B$}^{-1}_{k_2}\mbox{\boldmath $B$}^{-1}_u  =  (1+\rlap/u)
\frac{m_2+\rlap/p_2}{\sqrt{2m_2(k_2^0+m_2)}}
\end{equation}
and similar other formulae. (22) is obtained from the relation
$\mbox{\boldmath $B$}_u \rlap/x\mbox{\boldmath $B$}_u^{-1}
=   \gamma.\mbox{\boldmath $B$}_ux$,  which simply expresses that the
$\gamma^\mu$  matrices are forming a
4-vector, since from (20) we have
\begin{equation}
		(1+\gamma^0)\mbox{\boldmath $B$}^{-1}_{k_2}  =  (1+\gamma^0)
\frac{m_2+\rlap/k_2}{\sqrt{2m_2(k^0_2+m_2)}}
\end{equation}
and  $u = \mbox{\boldmath $B$}_u(1, \vec0)$,  $p_2 = \mbox{\boldmath
$B$}_uk_2$.  Finally, we use
\begin{equation}
		\mbox{\boldmath $B$}_u (\vec e.\vec\sigma) \gamma^0 \mbox{\boldmath
$B$}_u^{-1}  =  \gamma_5 \rlap/\epsilon_u,\qquad\qquad
\mbox{\boldmath $B$}_{u'} (\vec e^{\,\prime\displaystyle*}\!\!.\vec\sigma)
\gamma^0 \mbox{\boldmath $B$}_{u'}^{-1}  =
\gamma_5\rlap/\epsilon'^{\displaystyle*}_{u'}
\end{equation}

\section{Heavy mass limit: covariance and Isgur-Wise scaling.}

We now consider the heavy mass limit of (15, 16).  This limit is defined as
$m_1, m'_1 \to
\infty$  with  $v' = \frac{P'}{M'}$  and  $v = \frac PM$  fixed  ($M'$  and
$M$  are the final and
initial meson masses). It is also assumed that  $\frac M{m_1} \to 1$,
$\frac{M'}{m'_1} \to 1$.
It is then found that the integrand in (15) has a limit for fixed integration
variable  $\vec p_2$.
We have
\begin{eqnarray}
		\lefteqn{\frac{p_1}{m_1} \to v,}\hspace{1.5in} &&		\frac{p'_1}{m'_1} \to
v',\\
		\lefteqn{u \to v,}\hspace{1.5in} &&		u' \to v'\nonumber\\
		\lefteqn{\epsilon_u \to \epsilon_v = \epsilon,}\hspace{1.5in} &&
\epsilon'_{u'} \to
\epsilon'_{v'} = \epsilon'\nonumber\\
		\lefteqn{\frac{k_1^0}{m_1} \to 1,}\hspace{1.5in} &&		\frac{k_1'^0}{m'_1} \to
1\nonumber\\
		\lefteqn{k_2 \to \mbox{\boldmath $B$}_v^{-1}p_2,}\hspace{1.5in} &&		k'_2 \to
\mbox{\boldmath $B$}_{v'}^{-1}p_2\nonumber
\end{eqnarray}
Also,
\begin{equation}
		(\mbox{\boldmath $B$}_v^{-1}p_2)^0 = p_2.v,\qquad\qquad	(\mbox{\boldmath
$B$}_{v'}^{-1}p_2)^0  =  p_2.v'
\end{equation}
by invariance of the scalar product since  $\mbox{\boldmath $B$}_v^{-1}v =
(1,\vec0)$,  $\mbox{\boldmath $B$}_{v'}^{-1}v'
= (1,\vec0)$.  The limit of (15, 16)  is therefore
\begin{eqnarray}
	\lefteqn{\langle\vec P', \epsilon' | O |\vec P, \epsilon \rangle  =  \frac12
\frac1{\sqrt{v'^0v^0}} \int \frac{d\vec p_2}{(2\pi)^3}  \frac1{p^0_2}
\frac{\sqrt{(p_2.v')(p_2.v)}}{\sqrt{(p_2.v'+m_2)(p_2.v+m_2)}}}\qquad\\
			&&\frac14\;  \mathrm {
Tr}[O\gamma_5\rlap/\epsilon(1+\rlap/v)(\rlap/p_2+m_2)(1+\rlap/v')
\gamma_5\rlap/\epsilon'^{\displaystyle*}]
\; \varphi'(\overrightarrow{\mbox{\boldmath
$B$}_{v'}^{-1}p_2})^{\displaystyle*}  \varphi(\overrightarrow{\mbox{\boldmath
$B$}_v^{-1}p_2})
\nonumber
\end{eqnarray}
Now, this expression is explicitly a covariant function of the 4-vectors  $P'$
and  $P$  (or
$v'$  and  $v$)  because we have the integral with the invariant measure
$\frac{d\vec p_2}{p^0_2}$  of a covariant function of  $p_2$,  $v'$  and  $v$.
The only point
perhaps not immediately apparent is the invariance of
$\varphi'(\overrightarrow{\mbox{\boldmath
$B$}_{v'}^{-1}p_2})^{\displaystyle*}  \varphi(\overrightarrow{\mbox{\boldmath
$B$}_v^{-1}p_2})$.  In fact,
$\varphi'(\vec k)$  and  $\varphi(\vec k)$  being invariant by rotation are
function of $\vec k^2$
and, according to (26), we have
\begin{equation}
		(\overrightarrow{\mbox{\boldmath $B$}_{v'}^{-1}p_2})^2  =  (p_2.v')^2 -
m_2^2,\qquad\qquad
(\overrightarrow{\mbox{\boldmath $B$}_v^{-1}p_2})^2  =  (p_2.v)^2 - m_2^2
\end{equation}
Therefore  $\varphi'(\overrightarrow{\mbox{\boldmath
$B$}_{v'}^{-1}p_2})^{\displaystyle*}  \varphi(\overrightarrow{\mbox{\boldmath
$B$}_v^{-1}p_2})$ is a Lorentz scalar.

Using its covariance properties, eq. (27) can be reduced further. Indeed, (27)
can apparently be
expressed (for all Dirac matrices  $O$)  in term of three independent form
factors  $A$,  $B$,
$B'$,  defined by :
\begin{eqnarray}
	\lefteqn{A(v'.v)  =  \int \frac{d\vec p_2}{(2\pi)^3}  \frac1{p^0_2}
\frac{\sqrt{(p_2.v')(p_2.v)}}{\sqrt{(p_2.v'+m_2)(p_2.v+m_2)}}\;
\varphi'(\overrightarrow{\mbox{\boldmath
$B$}_{v'}^{-1}p_2})^{\displaystyle*}  \varphi(\overrightarrow{\mbox{\boldmath
$B$}_v^{-1}p_2})}\qquad\\
	\lefteqn{B'(v'.v)\, v'^\mu + B(v'.v)\, v^\mu}\qquad \nonumber\\
		&&=  \int \frac{d\vec p_2}{(2\pi)^3}  \frac1{p^0_2}
\frac{\sqrt{(p_2.v')(p_2.v)}}{\sqrt{(p_2.v'+m_2)(p_2.v+m_2)}}\; p_2^\mu\;\;
\varphi'(\overrightarrow{\mbox{\boldmath $B$}_{v'}^{-1}p_2})^{\displaystyle*}
\varphi(\overrightarrow{\mbox{\boldmath
$B$}_v^{-1}p_2}) \nonumber
\end{eqnarray}
After integration however, the expression
$(1+\rlap/v)(\rlap/p_2+m_2)(1+\rlap/v')$  becomes
	$$(1+\rlap/v)(B'\rlap/v'+B\rlap/v+m_2A)(1+\rlap/v')  =
(B+B'+m_2A)(1+\rlap/v)(1+\rlap/v')$$
 and, in fact, only the combination  $\xi = B+B'+m_2A$ appears. We obtain
finally the standard scaling formula :
\begin{equation}
	\langle\vec P', \epsilon' | O |\vec P, \epsilon \rangle  =  \frac12
\frac1{\sqrt{v'^0v^0}} \;
\frac14\; \mathrm { Tr}[O\gamma_5 \rlap/\epsilon(1+\rlap/v)
(1+\rlap/v')\gamma_5
\rlap/\epsilon'^{\displaystyle*}]\; \xi(v'.v)
\end{equation}
with the Isgur-Wise function  $\xi$  given by :
\begin{eqnarray}
		\lefteqn{\xi(v'.v)  =  \frac1{v'.v+1}  \int \frac{d\vec p_2}{(2\pi)^3}
\frac{\sqrt{(p_2.v')(p_2.v)}}{p^0_2}}\qquad\qquad\\
				&&\frac{p_2.(v'+v) + m_2(v'.v+1)}{\sqrt{(p_2.v'+m_2)(p_2.v+m_2)}}\;
\varphi'(\overrightarrow{\mbox{\boldmath $B$}_{v'}^{-1}p_2})^{\displaystyle*}
\varphi(\overrightarrow{\mbox{\boldmath
$B$}_v^{-1}p_2}) \nonumber
\end{eqnarray}
The matrix element with one or two pseudoscalar mesons are obtained from (30)
by omitting
$\gamma_5\rlap/\epsilon$  or  $\gamma_5\rlap/\epsilon'^{\displaystyle*}$  or
both under the trace.

One may notice that
\begin{equation}
		\xi(1)  =  \int \frac{d\vec p_2}{(2\pi)^3}\; \varphi'(\vec
p_2)^{\displaystyle*}
\varphi(\vec p_2)
\end{equation}
and that flavor independent forces and heavy mass limit entail  $\varphi' =
\varphi$,  so that
$\xi(1) = 1$.

Although they do not give a general expression like (27) and provide their
results only in a particular frame, by gathering the various
indications given in the paper by Close and Wambach \cite{close}, we apparently
 end up with precisely this
expression; we have also checked that their expression of the slope $\rho^2$,
given for harmonic oscillator wave functions, coincides with
ours. An important advantage of our approach is that we derive it from the
general
Bakamjian-Thomas formalism and we demonstrate the covariance and scaling
properties of the result.

Let us write the vector  ($O = V^\mu = \gamma^\mu$)  and axial  ($O = A^\mu =
\gamma^\mu\gamma_5$)
matrix elements with an initial pseudoscalar, following from (30) :
\begin{eqnarray}
	&&\sqrt{v'^0v^0}\; \langle\vec P'| V^\mu |\vec P\rangle   =   \frac12\;
(v'^\mu+v^\mu)\; \xi(v'.v)\\
	&&\sqrt{v'^0v^0}\; \langle\vec P', \epsilon' | V^\mu |\vec P\rangle  =  -
\frac i2\;
\sum_{\nu\rho\sigma}  \epsilon^{\mu\nu\rho\sigma} v'_\nu v_\rho
\epsilon'^{\displaystyle*}_\sigma\; \xi(v'.v)\nonumber\\
	&&\sqrt{v'^0v^0}\; \langle\vec P'| A^\mu |\vec P\rangle  =  0\nonumber\\
	&&\sqrt{v'^0v^0}\; \langle\vec P', \epsilon' | A^\mu |\vec P\rangle  =
\frac12\;  [(v'.v+1)\,
\epsilon'^{\mu\displaystyle*} - (v.\epsilon'^{\displaystyle*})\, v'^\mu]\;
\xi(v'.v)\nonumber
\end{eqnarray}
The corresponding form factors are
\begin{eqnarray}
	&&f_+(q^2)  =  V(q^2)  =  A_2(q^2)  =  A_0(q^2)  =  \frac{M'+M}{2\sqrt{M'M}}\;
 \xi(v'.v)	\\
	&&f_0(q^2)  =  A_1(q^2)  =  \frac{M'+M}{2\sqrt{M'M}}\;  [1 -
\frac{q^2}{(M'+M)^2}]\;  \xi(v'.v)
\nonumber\\
	&&(v'.v  =  \frac{M'^2+M^2-q^2}{2M'M}) \nonumber
\end{eqnarray}

\section{Lower bound on the $\rho^2$ slope.}

Let us now consider the slope  $\rho^2 = - \xi'(1)$  of the Isgur-Wise function
given by (31). We
establish here the following optimal lower bound of  $\rho^2$ :
\begin{equation}  \rho^2  > \frac34  \end{equation}
Eq. (31) is of the form
\begin{equation}
	\xi(v'.v)  =  \int \frac{d\vec p}{p^0}\;  F(p.v', p.v, v'.v)
\end{equation}
with
\begin{eqnarray}
	&&F(x', x, y)  =  \frac1{(2\pi)^3}\;  \frac{\sqrt{x'x}}{y+1} \;
\frac{x'+x+m_2(y+1)}{\sqrt{(x'+m_2)(x+m_2)}}\;
f(x'^2{-}m_2^2)^{\displaystyle*} f(x^2{-}m_2^2)
\quad\nonumber\\
	&&	f(\vec k^2) = \varphi(\vec k)
\end{eqnarray}
The integral in (36) depends only on the scalar product  $v'.v$  because the
integration measure
$d\vec p/p^0$  is Lorentz invariant. Expanding (36) around  $v'.v = 1$, we find
\begin{equation}
	\xi'(1)  =  - \int \frac{d\vec p}{p^0}\; [\frac13\,  \vec p^{\,2}\,
\partial_1\partial_2F(p^0, p^0,
1) - \partial_3F(p^0, p^0, 1)]
\end{equation}
And using eq. (37) for  $F$,  the following formula for the slope is obtained :
\begin{eqnarray}
	\lefteqn{\rho^2  =   \frac13  \int \frac{d\vec p}{(2\pi)^3}\,
[\vec\nabla\,p^0\varphi(\vec p)]^{\displaystyle*}.[\vec\nabla\,p^0\varphi(\vec
p)]}\qquad\qquad\\
				&&{}+  \int  \frac{d\vec p}{(2\pi)^3}\,
[\frac23+\frac14\,\frac{m^2_2}{(p^0)^2}-\frac13\,
\frac{m_2}{p^0{+}m_2}]\;\varphi(\vec p)^{\displaystyle*}\varphi(\vec
p)\nonumber
\end{eqnarray}
It is obvious that  $\rho^2 > 0$. In fact this expression is a positive
quadratic form of the wave
function  $\varphi$,  and the best lower bound of  $\rho^2$  is given by the
greatest lower bound
$B$  of the spectrum of the following corresponding self-adjoint operator  $T$:
\begin{equation}
	T  =  -  \frac13\:  p^0\, \Delta_{\vec p}\: p^0  +
\frac23+\frac14\,\frac{m^2_2}{(p^0)^2}-\frac13\,
\frac{m_2}{p^0{+}m_2}
\end{equation}
It is easily seen that the lower bound is obtained in the S-wave subspace.
Indeed, the reduction
$T_L$  of  $T$  to the $L$-orbital eigenspace, acting on the radial functions
$f(p)$  related to
$\varphi$  by $\varphi(\vec p)  =  |\vec p\,|^{-1}  f(|\vec p\,|) Y_L^M(\hat
p)$,  writes
\begin{equation}
	T_L  =  - \frac13\, p^0 \frac{d^2}{dp^2}\, p^0  +  \frac{L(L{+}1)}3\,
\frac{(p^0)^2}{p^2}  +
\frac23+\frac14\,\frac{m^2_2}{(p^0)^2}-\frac13\,\frac{m_2}{p^0{+}m_2}
\end{equation}
and, due to the "centrifugal barrier",  we have  $T_L \le T_{L'}$  if  $L <
L'$.

We may therefore concentrate on  (41)  with  $L = 0$.  Numerical integration of
the ordinary
differential equation  $T_0f = B'f$  (with initial condition  $f(0) = 0$,
$f'(0) = 1$)  gives a
strong indication that the lower bound of the spectrum is
\begin{equation}
		B = \frac34.
\end{equation}
Indeed, one finds that every $B' \ge \frac34$  is a generalized eigenvalue and
that the (non
normalisable) eigenfunction is oscillating for  $B' > \frac34$   and non
oscillating for  $B' =
\frac34$.

For a proof of (42), we use a {\it unitary} transformation  $U : \mathrm {
L}^2([0,\infty[) \to
\mathrm { L}^2([0,\infty[)$  which has the virtue of converting  $T_L$  into an
ordinary Schr\"odinger
operator. $U$  is just the following change of variable :
\begin{eqnarray}
	&&(Uf)(x)  =  \sqrt{m_2\,\mathrm { cosh}(x)}\: f(m_2\,\mathrm { sinh}(x)),\\
	&&(U^{-1}f)(p)  =  \frac1{\sqrt{p^0}}\,  f(\mathrm { Arccosh} \frac
p{m_2})\nonumber
\end{eqnarray}
The operator $p^0 d/dp$  becomes :
\begin{equation}
	U\, p^0 \frac d{dp}\, U^{-1}  =  \frac d{dx}  - \frac12\,  \mathrm { tanh}(x)
\end{equation}
and the operator  $T_0$  (eq. (41) with $L = 0$)  becomes
\begin{equation}
	U\, T_0\, U^{-1}  =  \frac34  +  \frac13\,  (- \frac{d^2}{dx^2}  -
\frac1{1+\mathrm { cosh}(x)})
\end{equation}
Here we can see that  $B \le \frac34$   because the (always positive)
expectation value of the operator
$-\,d^2/dx^2$   on a spreading function  $f_\lambda(x)  =  \sqrt\lambda
f(\lambda x)$,  $\lambda \to
0$,  ($||f_\lambda|| = ||f|| = 1$)  goes to  $0$.  On the other hand we have
$B \ge  \frac34$
because the operator between parenthesis in (45) is positive, as can be seen
from the following
identity:
\begin{equation}
	- \frac{d^2}{dx^2}  -  \frac1{1+\mathrm { cosh}(x)}  =  (\frac d{dx} +
\frac1{\mathrm { sinh}(x)}) (-\frac d{dx} +
\frac1{\mathrm { sinh}(x)})
\end{equation}

Finally, one sees as follows that the lower bound  $B=3/4$  of $\rho^2$  cannot
be attained.  If
$\rho^2 = 3/4$,  the expectation value of  (46)  vanishes on the normalized
function  $f(x)$  which
corresponds to $\varphi(\vec p)$. This implies  $(- \frac d{dx} +
\frac1{\mathrm { sinh}(x)}) f(x) = 0$.
However, the solutions  $f(x) = c\;\mathrm { tanh}(\frac x2)$  of this equation
are not normalizable.

The bound (35) is obviously in agreement with Bjorken's lower bound
\cite{bjorken} $\rho^2 >1/4$. It does not contradict Voloshin's upper bound
\cite{voloshin} $\rho^2 < 0.75 \pm 0.15$, but not much room is left and a
careful study of the relativistic generalisation of Thomas-Reiche-Kuhn sum rule
in our model should be performed. Neither does it contradict de Rafael-Taron's
\cite{derafael} conservative estimate $\rho^2 < 1.7$ not to speak of their
rigorous bound $\rho^2 < 6.0$.

Let us emphasize that the bound $B=3/4$ applies to a large class of quark
models, which at
least in this heavy quark limit show very remarkable properties, and is
independent of particular parameters of the models. Whence the interest of
testing it experimentally. But this is not an easy task. Let us recall that
$\rho^2$ is not directly comparable to the $\hat \rho^2$ measured by CLEO II
experiment \cite{CLEO}, from which it should differ by QCD radiative
corrections and $1/m_c$ corrections \cite{neubert}. The QCD radiative
corrections enhance  $\hat \rho^2$ versus $\rho^2$; the $1/m_c$ corrections are
not known from exact QCD, and cannot be safely deduced from our model in which
they are not covariant. The central value of the present CLEO II data: $\hat
\rho^2=0.87 \pm 0.12 \pm 0.08$ \cite{CLEO} is compatible with our lower bound,
although not far from it. But one must keep in mind the above-mentioned
corrections. A further improvement  of data may prove very instructive.

 The wave functions that saturate the lower bound are, for example, of the
type:
\begin{equation}
\varphi(\vec p) \propto \sqrt\epsilon\;(p^0)^{-(1.5+\epsilon)}
\end{equation}
for which $\rho^2 \to 0.75$ when $\epsilon \to 0$.

Harmonic oscillator wave functions,
\begin{equation}
\varphi(\vec p)= (2\pi)^{\frac32}\left({R/ \sqrt{\pi} } \right)^{\frac 3 2}
e^{-R^2\vec p^2 /2},\end{equation}

which are most commonly used, yield $\rho^2$ well above the bound 0.75. Indeed
we found that the lower bound for harmonic oscillator wave functions is
$1.208$. In the limit $m^2 R^2=0$ they give $\rho^2 = 5/4$. The minimum 1.208
is obtained for $m^2 R^2 \simeq 0.05$ which is far below the standard quark
model parameters, say, $m^2 \simeq 0.1$ GeV$^2$, $R^2 \simeq 6$ GeV$^{-2}$,
leading to $m^2 R^2 \simeq 0.6$. Using the latter values we obtain $\rho^2
\simeq 1.37$ which is, as expected, very similar to Close and Wambach's
\cite{close}: $\rho^2 \simeq (1.19)^2$.
In the weak coupling limit $m^2 R^2 \to \infty$ one gets
\begin{equation} \rho^2 = \frac{m^2 R^2}2 + 1 + O\left(\frac 1 {m^2
R^2}\right)\end{equation}

and the Isgur-Wise function takes the very simple form in the large $m^2 R^2$:

\begin{equation} \xi(v.v')=\frac {2}{1+v.v'} \frac 1 {\sqrt{v.v'}} \exp[-m^2
R^2 (v.v'-1)/2]\left (1 + O\left(\frac 1 {m^2 R^2}\right)\right)
\end{equation}

We have checked numerically that expression (50) is a surprisingly good
approximation to the Isgur-Wise function obtained with the standard quark model
 parameters: $m^2 \simeq 0.1$ GeV$^2$, $R^2 \simeq 6$ GeV$^{-2}$, although $m^2
R^2$ does not look so large. this amusingly simple formula is not
phenomenologically very  useful since, as already stressed, the harmonic
oscillator wave functions yield a $\rho^2$  much above the lower bound and
above experiment.

\section*{Acknowledgements.}

This work was supported in part by the CEC Science Project
SC1-CT91-0729 and by
the Human Capital
and Mobility Programme, contract CHRX-CT93-0132.

\end{document}